\documentclass[runningheads]{svmult}

\usepackage{makeidx}   
\usepackage{graphicx}  
\usepackage{subeqnar}  
\usepackage{multicol}  
\usepackage{physprbb}  
\makeindex             

\usepackage{amsmath,amssymb,amsfonts}  

\begin{document}
\title*{Asymptotically flat and regular Cauchy data}

\toctitle{Asymptotically flat and regular Cauchy data}
%
%
\titlerunning{Asymptotically flat and regular Cauchy data}
%
\author{Sergio Dain}
\authorrunning{S. Dain}
%
%
\institute{Max-Planck-Institut f\"ur Gravitationsphysik\\
Am M\"uhlenberg 1\\
14476 Golm\\
Germany  }

\maketitle              

\begin{abstract}
  I describe the construction of a large class of asymptotically flat
  initial data with non-vanishing mass and angular momentum for which
  the metric and the extrinsic curvature have asymptotic expansions at
  space-like infinity in terms of powers of a radial coordinate. I
  emphasize the motivations and the main ideas behind the proofs.
\end{abstract}

\newpage

\section{Introduction}

Suppose we want to describe an isolated self-gravitating system.  For
example a star, a binary system, a black hole or colliding  black
holes.  Typically these astrophysical systems are
located far away from the Earth, so that we can receive from
them only   electromagnetic and gravitational radiation. 
How is this radiation? For example one can ask  how much energy is
radiated, or which are the typical frequencies for some systems. This
is the general problem we want to study. 
These systems are expected to be described by the Einstein field
equations.  It is in principle possible to measure this radiation and
compare the results with the predictions of the equations. 

There are no explicit solution to the Einstein equations that can
describe such systems. Since the equations are very complicated it is
hard to believe that such explicit solution can ever be founded.
Instead of trying to solve the complete equations at once, we use the
so called 3+1 decomposition.  We split the equations into
``constraint'' and ``evolution''. First we give appropriate initial
data: that is, a solution of the constraint equations.  Once we have
chosen the initial data, the problem is completely fixed. Then, we use
the evolution equations to calculate the whole space-time. From the
evolution we can compute physical relevant quantities like the wave
form of the emitted gravitational wave.  This method of solving the
equations is consistent with the idea that in physic we want to make
predictions, that is: knowing the system at a given time we want to
predict its behavior in the future.  It is of course in general very
hard to compute the evolution of the data, one has to use numerical
techniques.  The question we want to study here is: what are the
appropriate initial data for an isolated system?

We can think of initial data for Einstein's equation as given a
picture of the space time at a given time. It consists of a three
dimensional manifold $\tilde S$ with some fields on it. The fields
must satisfy the constraint equations. If the data describe an
isolated system, the manifold $\tilde S$ is naturally divided in two
regions. One compact region $\Omega$ which ``contains the sources'',
and its complement $\tilde S \setminus \Omega$ which is unbounded. We
will call the latter one the asymptotic region. Of course, $\Omega$
can be as large as we want, the only requirement is that it be
bounded. The fact that $\tilde S \setminus \Omega$ is unbounded means
that we can go as far a we want from the source region $\Omega$, this
capture the idea that the sources in $\Omega$ are isolated.  That
``the sources are in $\Omega$'' suggest that the field decays in
$\tilde S \setminus \Omega$. Then, the initial data approach flat
initial data near infinity. We call them \emph{asymptotically flat
  initial data}. An opposite situation is   when we want to study the
universe as a whole. In this case one usually consider initial data
where the manifold $\tilde S$ is compact without boundary. 
\begin{figure}[htbp]
  \begin{center}
    \includegraphics[width=6cm]{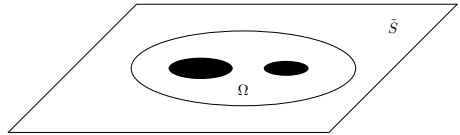}
    \caption{Initial data for two stars. The region $\Omega$ contains
      the matter sources, which are represented by black regions. The
      manifold $\tilde S$ is $\mathbb R^3$.}
    \label{fig:sources}
  \end{center}
\end{figure}

The simplest and most important example of asymptotically flat initial
data is the case when $\tilde S$ is $\mathbb{R}^3$ and $\Omega$ some
ball. Such data can describe, for example, ordinary stars. Consider an
initial data set for a binary system, as is shown in Fig.
\ref{fig:sources}. The stars are inside $\Omega$.  Ordinary stars emit
light, then in the asymptotic region we will have electromagnetic
field beside the gravitational one.  We can also have some dilute gas
in this region. It is important to recall that we do not require
vacuum in the asymptotic region. We only require that the field decays
properly outside $\Omega$.  However, in many situations it is a good
approximation to assume vacuum in that region.  When they evolve, the
stars will follow some trajectory. Presumably an spiral orbit. The
radius of the orbit will decrease with time, since the system loses
energy in the form of gravitational waves.  At late times, the system
will settle down to a final stationary regime. This final state can be
a rotating, stationary star. Or, if the initial data have enough mass
concentrated in a small $\Omega$, a spinning black hole. We show the
conformal diagrams of this two cases in Fig. \ref{fig:conf1} and Fig.
\ref{fig:conf2}.  The gravitational radiation is measured at null
infinity, where the observer is placed.

\begin{figure}
\begin{minipage}[t]{0.45\linewidth}
  \centering
   \includegraphics[width=3cm]{evolution.1}
    \caption{Conformal diagram of the space time generated by the
      initial data given in Fig. \ref{fig:sources}, in the case in
      which the final state is a stationary star.}
    \label{fig:conf1}
  \end{minipage}
\hspace{0.1\linewidth}
\begin{minipage}[t]{0.45\linewidth}
  \centering
     \includegraphics[width=3cm]{evolutbh.1}
    \caption{Conformal diagram of the space time generated by the
      initial data given in Fig. \ref{fig:sources}, in the case in
      which the final state is a black hole.}
    \label{fig:conf2}
  \end{minipage}
\end{figure}

\begin{figure}
\begin{minipage}[t]{0.45\linewidth}
  \centering
     \includegraphics[width=5cm]{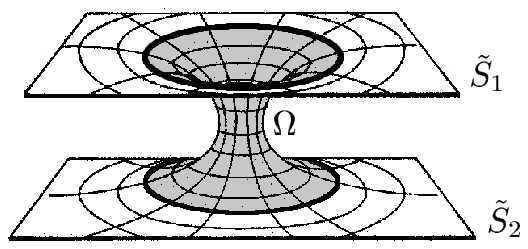}
    \caption{One black hole initial data. The asymptotic region has
      two disjoint components $\tilde S_1$ and $\tilde S_2$. The
      compact set $\Omega$ is represented by the shadowed region.}
    \label{fig:osch}
  \end{minipage}
\hspace{0.1\linewidth}
\begin{minipage}[t]{0.45\linewidth}
  \centering
   \includegraphics[width=5cm]{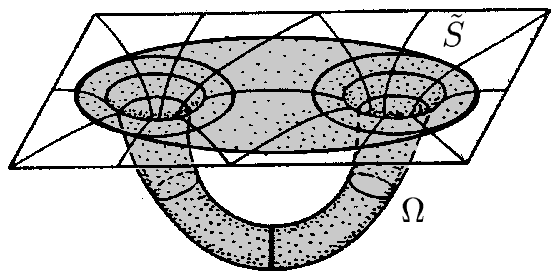}  
    \caption{Initial data for a Misner wormhole \cite{Misner60}.  The
      asymptotic region has only one component. The compact set
      $\Omega$ is represented by the shadowed region.}
    \label{fig:wormhole}
    \end{minipage}
\end{figure}

\begin{figure}
\begin{minipage}[t]{0.45\linewidth}
  \centering
     \includegraphics[width=4.5cm]{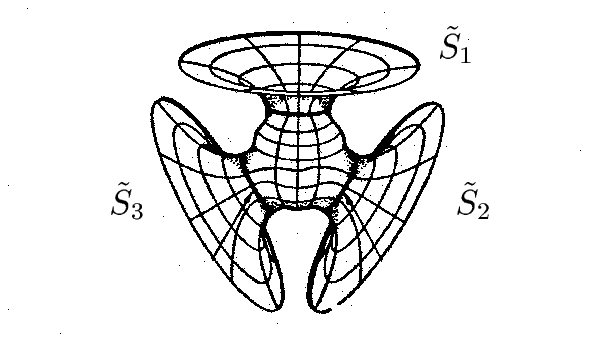}
    \caption{The Brill-Lindquist initial data for two black holes
      \cite{Brill63}. The 
      data have three  asymptotic ends.}
\label{fig:bl}
  \end{minipage}
\hspace{0.1\linewidth}
\begin{minipage}[t]{0.45\linewidth}
  \centering
   \includegraphics[width=2cm]{sph.2}  
    \caption{A compactification of these data in the sphere.}
    \label{fig:cbl}
    \end{minipage}
\end{figure}

\begin{figure}
\begin{minipage}[t]{0.45\linewidth}
  \centering
     \includegraphics[width=4cm]{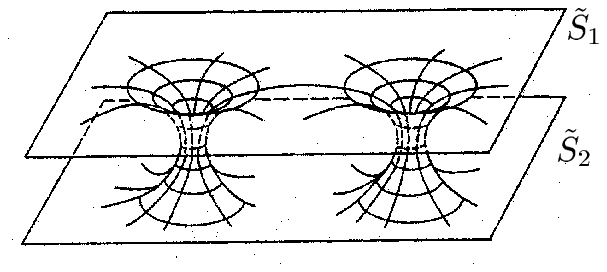}
    \caption{The Misner initial data for two black holes \cite{Misner63}. The 
      data have three  asymptotic ends.}
\label{fig:misner}
  \end{minipage}
\hspace{0.1\linewidth}
\begin{minipage}[t]{0.45\linewidth}
  \centering
   \includegraphics[width=3cm]{torus.2}
    \caption{A compactification of these data in the torus.}
    \label{fig:cmisner}
    \end{minipage}
\end{figure}

Since the sources are in $\Omega$, the fields in this region can be
very complicated. They depend on the kind of matter that forms the
stars.  Remarkable, even the topology of $\Omega$ can be complicated.
The example in Fig. \ref{fig:sources} has trivial topology, however
non trivial topologies are relevant for black hole initial data.
Consider the initial data for the Schwarzschild and Kerr black hole.
In this case $\tilde S=S^2\times \mathbb{R}$ and
$\Omega=S^2\times[-a,a]$ for some constant $a$, as it is shown in Fig.
\ref{fig:osch}.  The asymptotic region of these data has two
disconnected components $\tilde S\setminus \Omega=\tilde S_1+ \tilde
S_2$. Each component is diffeomorphic to $\mathbb{R}^3$ minus a
compact ball. We say that the data have two asymptotic ends. That is,
there exist two disconnected ``infinities''. Observations can be made
in one of them but not in both. It is not clear the physical relevance
of this extra asymptotic end. A normal collapse, as is shown in Fig.
\ref{fig:conf2}, will not have it. Therefore, one can think that only
one asymptotic end is astrophysically relevant, being the other one
just a mathematical peculiarity.  The topology is not fixed by the
Einstein equations. In this sense it remains quite arbitrary, one can
thing that it is a kind of boundary conditions that has to be extra
imposed to the equations.  However, non trivial topologies appear
naturally in the study of vacuum stationary black holes. A space time
is stationary if it admits a timelike Killing vector field. The
Schwarzschild and Kerr metrics are stationary. One can prove that
every vacuum stationary asymptotically flat space time with trivial
topology must be Minkowski. That is the non trivial topology is the
``source'' of the gravitational field in the vacuum stationary space
times.  In Fig.  \ref{fig:wormhole} we show another example with
different a topology, in this case we have only one asymptotic end.
Other topologies with many asymptotic ends are of interest because
their evolution in time may describe the collision of several black
holes.  In Fig.  \ref{fig:bl} and Fig.  \ref{fig:misner} we show some
examples.

One usually has the idea that the matter sources generate
 gravitational field.  To some extend this is true for  many cases of
astrophysical interest. However, as we have seen, a pure vacuum
initial data with non trivial topologies can have mass. Moreover, one
can also have a pure gravitational radiation  initial data, that is
a  vacuum initial data with trivial topology with   non zero
mass. These kind of data are important in order to understand
properties of the radiation itself which do not depend on the specific
matter models. It is even possible to form a black hole with these
type of data (see \cite{Beig91c}).

In contrast to $\Omega$, the asymptotic region is very simple. The
fields on it are approximately flat and its topology is $\mathbb R^3$
minus a compact ball. We want to analyze the fields in this region.
It is of course true that the fields there are determined by the
fields in $\Omega$. But some important properties of them do not
depend very much on the detailed structure of the sources in $\Omega$.
The most important of these properties is that the initial data will
have a positive mass if the sources satisfies some energy condition.
In this article we want to study some other properties of the fall of the
fields in the asymptotic region which will be also independent on the
 sources in $\Omega$. It is important to note that we can only
measure precisely this kind of properties, because we can not prepare
the initial conditions of an astrophysical system in the laboratory
and then it is impossible to find the exact initial data for a real
system.  We can only analyze those effects which do not depend very
much on fine details of the sources.

In order to simplify the hypothesis of the theorems, we will assume in
this article that there are no matter sources in $\tilde S$.  All the
results will remain valid under suitable assumptions on the matters
sources.

We  summarize the discussion above in the following definition.
An \emph{initial data set} for the Einstein vacuum equations is given by a
triple  $(\tilde S, \tilde h_{ab}, \tilde \Psi_{ab})$  where $\tilde S$
is a connected 3-dimensional manifold, $\tilde h_{ab} $ a (positive
definite) Riemannian metric, and $\tilde \Psi_{ab}$ a symmetric tensor
field on 
$\tilde S$. They satisfy the vacuum constraint equations
\begin{equation}
 \label{const1}
\tilde D^b \tilde \Psi_{ab} -\tilde D_a \tilde \Psi=0,
\end{equation}
\begin{equation}
 \label{const2}
\tilde R + \tilde \Psi^2-\tilde \Psi_{ab} \tilde \Psi^{ab}=0,
\end{equation}
on $\tilde S$, where $\tilde D_a$ is the covariant derivative with
respect to $\tilde h_{ab}$, $\tilde R$ is the trace of the
corresponding Ricci tensor, and $\tilde \Psi=\tilde h^{ab} \tilde
\Psi_{ab}$.  The data will be called \emph{asymptotically flat} with
$N$ \emph{asymptotic ends}, if for some compact set $\Omega$ we have
that  $\tilde
S\setminus \Omega =\sum_{k=1}^N\tilde S_k$, where $\tilde S_k$ are
open sets such that each $\tilde S_k$ can be mapped by a coordinate
system $\tilde x^j$ diffeomorphically onto the complement of a closed
ball in $\mathbb{R}^3$ such that we have in these coordinates
\begin{equation} 
\label{pf1}
\tilde h_{ij}=(1+\frac{2m}{\tilde r})\delta_{ij}+O(\tilde r^{-2}),
\end{equation}
\begin{equation} 
\label{pf2}
\tilde \Psi_{ij}=O(\tilde r^{-2}),
\end{equation}
as $\tilde r= ( \sum_{j=1}^3 ({\tilde x^j})^2 ) ^{1/2} \to \infty$ in
each set $\tilde S_k$.
Here the constant $m$ denotes the mass of the data, $a,b,c...$ denote
abstract indices, $i,j,k...$, which take values $1, 2, 3$, denote
coordinates indices while $\delta_{ij}$ denotes the flat metric with respect
to the given coordinate system $\tilde x^j$. Tensor indices will be moved
with the metric $h_{ab}$ and its inverse $h^{ab}$. We set $x_i = x^i$ and
$\partial^i = \partial_i$. These  conditions guarantee that the mass, the
momentum, and the angular momentum of the initial data set are well
defined in every end.

We want to analyze the higher order terms in (\ref{pf1}) and
(\ref{pf2}). 
  For example, the terms
$O(\tilde r^{-2})$  could have the form
$\tilde r^{-2}\sin(\tilde r^{10})$. This function is certainly $O(\tilde
r^{-2})$, but any derivative of it will blow up at infinity. Should such
complicated terms be admissible in a description of a  realistic
isolated systems?  In  Newton's theory and  Electromagnetism one
can give a definitive answer to this question: the fields have a fall
off behavior like powers in a radial coordinate. Take for example a
matter density $\rho$ with compact support. The Newton's gravitational
potential $u$ satisfies the Poisson equation
\begin{equation}
  \label{eq:2}
  \Delta u =4\pi \rho,
\end{equation}
where $\Delta$ is the Laplacian. Outside the support of $\rho$ the
potential $u$ satisfies $\Delta u=0$. It is a well known  that
every harmonic function that  goes to zero at infinity has an expansion
of the form 
\begin{equation}
  \label{eq:3}
  u=\sum_{k \geq 1} 
\frac{ u^{k-1}}{\tilde r^k},
\end{equation}
where $u^k$ are the spherical harmonics of order $k$.  In
this case the field equations force the potential  to have the  fall off
behavior \eqref{eq:3}.

The situation for the Einstein equation is  more
complicate.  
In  analogy with \eqref{eq:3}, one can ask the question whether there
exist a class of initial data such that 
the metric and the extrinsic curvature admit near
space-like infinity asymptotic expansions of the form
\begin{equation} 
\label{he1}
\tilde h_{ij}\sim (1+\frac{2m}{\tilde r})\delta_{ij}+\sum_{k \geq2} 
\frac{\tilde h^k_{ij}}{\tilde r^k},
\end{equation}
\begin{equation} 
\label{Psie2}
\tilde \Psi_{ij}\sim \sum_{k \geq2} \frac{\tilde \Psi^k_{ij}}{\tilde r^k},
\end{equation}
where $\tilde h^k_{ij}$ and $\tilde \Psi^k_{ij}$ are smooth function
on the unit 2-sphere (thought as being pulled back to the spheres
$\tilde{r} = const.$ under the map $\tilde{x}^j \rightarrow
\tilde{x}^j/\tilde{r}$). In this article I want to give an answer to
this question. It is not only for convenience or aesthetic reasons
that we want to avoid terms like $\tilde r^{-2}\sin(\tilde r^{10})$ in
the expansions.  They are also very difficult to handle numerically.

In order to see how the gravitational field behaves near infinity, it
is natural to study first some examples.  Consider asymptotically flat
static space-times. We say that a space-time is static if it admits a
hypersurface orthogonal time like Killing vector. One can take one of
this hypersurfaces and analyze the fields $\tilde h_{ab}$ and $\tilde
\Psi_{ab}$ on it. The simplest static space time is the Schwarzschild
metric. In this case we have $\tilde h_{ij}=(1+m/(2\tilde
r))^4\delta_{ij}$, $\tilde \Psi_{ij}=0$, which is certainly of the
form (\ref{he1}), (\ref{Psie2}). For general static, asymptotically
flat space times it can be proved that the initial data have also
asymptotic expansions of the form (\ref{he1}) and (\ref{Psie2}).
Moreover, in this case the fields are analytic functions of the
coordinates. Then the expansions (\ref{he1}) and (\ref{Psie2}) are in
fact convergent powers series, in analogy with (\ref{eq:3}).  This
result is far from being obvious. It was proved by Beig and Simon in
\cite{Beig80} based on an early work by Geroch \cite{Geroch70}.

One can go a step further and consider asymptotically flat stationary
solutions. A stationary space time admits a timelike Killing field
which is in general non hypersurface orthogonal.  These space times
describe rotating stars in equilibrium.  The twist of the Killing
vector is related to the angular velocity of the star. In this case
there are no preferred hypersurfaces.  The most important example of
stationary solution is the Kerr metric.  It can be seen that for
$t=const.$ slice of the Kerr metric in the standard Boyer-Lindquist
coordinates the fields $\tilde h_{ab}$ and $\tilde \Psi_{ab}$ satisfy
(\ref{he1}) and (\ref{Psie2}).  For general stationary asymptotically
flat solutions there also exist slices such that (\ref{he1}) and
(\ref{Psie2}) holds.  The essential part of this result was also
proved by Beig and Simon in \cite{Beig81} (see also \cite{Kundu81}).
However, in these works  the expansions are made in the abstract manifold
 of the orbits of the Killing vector field. In
contrast with the static case this manifold does not correspond to any
hypersurface of the space time.  One has to translate this result in
terms of the metric and the extrinsic curvature of some slices. This
last step was made in \cite{Dain01b}.  As in the static case, the
fields are analytic functions of the coordinates.

We have shown that there exist non trivial examples of initial data
which satisfies \eqref{he1} and \eqref{Psie2}. But how general are
these expansions? For example, is it possible to have data with
gravitational radiation that satisfies \eqref{he1} and \eqref{Psie2}?
I want to show that in fact there exists a large class of
asymptotically flat initial data which have the asymptotic behavior
\eqref{he1} and \eqref{Psie2}. These data will not be, in general,
stationary.

The interest in such data is twofold. First,  the evolution of
asymptotically flat initial data near space-like and null infinity has
been studied in considerable detail in \cite{Friedrich98}. In that article
has been derived in particular a certain ``regularity condition'' on the
data near space-like infinity, which is expected to provide a criterion
for the existence of a smooth asymptotic structure at null infinity. To
simplify the lengthy calculations, the data considered in
\cite{Friedrich98} have been assumed to be time-symmetric and to admit a
smooth conformal compactification. With these assumptions the regularity
condition is given by a surprisingly succinct expression.  With the
present work we  want to provide data which will allow us to perform the
analysis of
\cite{Friedrich98} without the assumption of time symmetry but which are
still ``simple'' enough to simplify the work of generalizing the regularity
condition to the case of non-trivial second fundamental form. 
Second, the ``regular finite initial value problem near space-like infinity'',
formulated and analyzed in \cite{Friedrich98}, suggests how to calculate
numerically entire asymptotically flat solutions to Einstein's vacuum field
equations on finite grids. In the present article I  provide data for such
numerical calculations which should allow us to study interesting
situations while keeping a certain simplicity in the handling of the
initial data.   

The difficulty of constructing data with the asymptotic behavior
(\ref{he1}), (\ref{Psie2}) arises from the fact that the fields need
to satisfy the constraint equations (\ref{const1}) and (\ref{const2}).
Part of the data, the ``free data'', can be given such that they are
compatible with (\ref{he1}), (\ref{Psie2}). However, the remaining
data are governed by elliptic equations (the constraint equations will
reduce to elliptic equations as we will see below) and we have to show
that (\ref{he1}), (\ref{Psie2}) are in fact a consequence of the
equations and the way the free data have been prescribed.

To employ the standard techniques to provide solutions to the
constraints, we assume   
\begin{equation}
\label{physmax}
\tilde \Psi = 0, 
\end{equation}
such that the data correspond to a hypersurface which is maximal in the
solution space-time.

I give an outline of the  results which are available so far (see
\cite{Dain99} and \cite{Dain01}  for more details and proofs).
Because of the applications indicated above, we wish to control in
detail the conformal structure of the data near space-like infinity.
Therefore we shall analyze the data in terms of the conformal
compactification $(S, h_{ab}, \Psi_{ab})$ of the ``physical''
asymptotically flat data. Here $S$ denotes a smooth, connected,
orientable, compact 3-manifold. Take an arbitrary point $i$ in $S$ and
define $\tilde{S} = S\backslash \{ i \} $.  The point $i$ will
represent, in a sense described in detail below, space-like infinity
for the physical initial data.  The physical manifold $\tilde S$ is
essentially the stereographic projection of the compact manifold $S$.
For example if we chose $S$ to be $S^3$ then $\tilde S$ will be
$\mathbb R^3$. 

\begin{figure}
\begin{center}
\includegraphics[width=6cm]{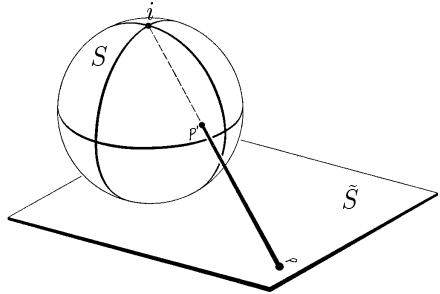}

\caption{The stereographic projection of $S^3$ in $\mathbb R^3$. A
  neighborhood $B_a$ of the point $i$ is mapped into the asymptotic
  region of the physical manifold $\tilde S$.}
\end{center}
\end{figure}

\begin{figure}
\begin{minipage}[t]{0.45\linewidth}
  \centering
     \includegraphics[width=4cm]{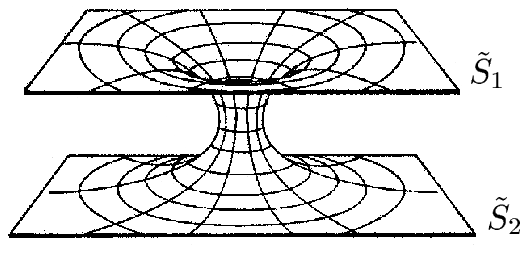}
    \caption{Initial data with two asymptotic ends.}
\label{fig:sch}
  \end{minipage}
\hspace{0.1\linewidth}
\begin{minipage}[t]{0.45\linewidth}
  \centering
   \includegraphics[width=2cm]{sph.1}  
    \caption{A compactification of these data in the sphere.}
    \label{fig:csch}
    \end{minipage}
\end{figure}

\begin{figure}
\begin{minipage}[t]{0.45\linewidth}
  \centering
     \includegraphics[width=4cm]{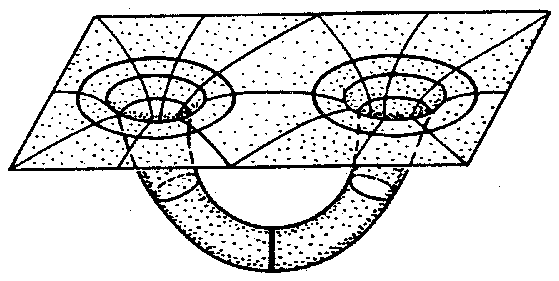}
    \caption{The Misner wormhole initial data. The 
      data have only one asymptotic end.}
\label{fig:ww}
  \end{minipage}
\hspace{0.1\linewidth}
\begin{minipage}[t]{0.45\linewidth}
  \centering
   \includegraphics[width=3cm]{torus.1}
    \caption{A compactification of these data in the torus.}
    \label{fig:cww}
    \end{minipage}
\end{figure}

Working with $S$ and not with $\tilde S$, has several technical
advantage. It is simpler to prove existence of solutions for an
elliptic equation on a compact manifold than in a non compact one. The
price that we have to pay is that the equations will be singular at
$i$. However this singularity is mild. It is also simpler to analyze
the fields in terms of local differentiability in a neighborhood of
$i$ than in terms of fall off expansions at infinity in $\tilde S$.
Moreover, this technique is also useful to construct initial data with
non-trivial topology.  By singling out more points in $S$ and by
treating the fields near these points in the same way as near $i$ we
could construct data with several asymptotically flat ends. In Fig.
\ref{fig:sch} -- \ref{fig:ww} we show some examples.  All the
following arguments equally apply to such situations, however, for
convenience we restrict ourselves to the case of a single
asymptotically flat end.

We assume that $h_{ab}$ is a positive definite metric on $S$ with covariant
derivative $D_a$ and $\Psi_{ab}$ is a symmetric tensor field which is
smooth on $\tilde{S}$. In agreement with (\ref{physmax}) we shall assume
that
$\Psi_{ab}$ is trace free, 
\[
h^{ab}\,\Psi_{ab} = 0. 
\]
The fields above are related to the physical fields by rescaling
\begin{equation}
\label{hPrescale}
\tilde{h}_{ab} = \theta^4\,h_{ab},\,\,\,\,\,\,\, 
\tilde{\Psi}_{ab} = \theta^{-2}\,\Psi_{ab},
\end{equation}
with a conformal factor $\theta$ which is positive on $\tilde{S}$.
For the physical fields to satisfy the vacuum constraints we need to assume
that
\begin{equation} 
\label{diver}
D^a \Psi_{ab}=0 \quad\mbox{on}\quad \tilde{S},
\end{equation}
\begin{equation} 
\label{Lich}
(D_{b}D^{b}-\frac{1}{8}R)\theta=-\frac{1}{8}\Psi_{ab}\Psi^{ab}\theta^{-7}
\quad\mbox{on}\quad \tilde{S}. 
\end{equation}
Equation (\ref{Lich}) for the conformal factor $\theta$ is the Lichnerowicz
equation, transferred to our context. 

Let $x^j$ be $h$-normal coordinates centered at $i$ such that 
$h_{kl}= \delta_{kl}$ at $i$ and set $r = (\sum_{i=1}^3 (x^j)^2 ) ^{1/2}$. 
To ensure asymptotical flatness of the data (\ref{hPrescale}) we require   
\begin{equation} 
\label{Psii}
\Psi_{ab} = O(r^{-4}) \quad\mbox{as}\quad r \rightarrow 0,  
\end{equation}
\begin{equation} 
\label{thetai}
\lim_{r\to 0}r\theta = 1.
\end{equation}
In the coordinates $\tilde{x}^j = x^j/r^2$ the fields (\ref{hPrescale})
will then satisfy (\ref{pf1}), (\ref{pf2}) (cf. \cite{Friedrich88},
\cite{Friedrich98} for this procedure).

Not all data as given by (\ref{hPrescale}), which are derived from data
$h_{ab}$, $\Psi_{ab}$ as described above, will satisfy conditions
(\ref{he1}), (\ref{Psie2}).  We will have to impose extra conditions and
we want to keep these conditions as simple as possible.  

Since we assume the metric $h_{ab}$ to be smooth on $S$, it will
only depend on the behavior of $\theta$ near $i$ whether condition
(\ref{he1}) will be satisfied. Via equation (\ref{Lich}) this behavior
depends on $\Psi_{ab}$.  What kind of condition do we have to impose on
$\Psi_{ab}$ in order to achieve (\ref{he1}) ?  

The following space of functions will play an important role in our
discussion. Denote by $B_a$ the open ball with
center $i$ and radius $r = a > 0$, where $a$ is chosen small enough such
that $B_a$ is a convex normal neighborhood of $i$.   
A function $f\in C^\infty(\tilde S)$ is said to be in $E^\infty(B_a)$
if on  $B_a$ we can write $f = f_1 +rf_2 $
with $f_1, f_2 \in C^\infty(B_a)$ (cf. definition \ref{dEm}).
An  answer to our question is given by following theorem: 

\begin{theorem} 
\label{T1}
Let $h_{ab}$ be a smooth metric on $S$ with positive Ricci scalar $R$. 
Assume  that $\Psi_{ab}$ is smooth in $\tilde S$ and satisfies on
$B_a$   
\begin{equation} 
\label{condd}
r^8\Psi_{ab} \Psi^{ab}\in E^\infty(B_a).
\end{equation} 
Then there exists  on $\tilde S$ a unique solution $\theta$ of 
equation (\ref{Lich}), which is positive, satisfies (\ref{thetai}), 
and has in $B_a$ the form
\begin{equation} 
\label{1f}
\theta = \frac{\hat \theta}{r}, \quad \hat \theta \in E^\infty(B_a),
\quad \hat \theta (i)=1.
\end{equation}
\end{theorem}

In fact, we will get slightly more detailed information. We find that 
$\hat \theta= u_1 + r\,u_2$ on $B_a$ with $u_2 \in E^\infty(B_a)$ and a
function $u_1\in C^\infty(B_a)$ which satisfies $u_1=1+O(r^2)$ and
\[
(D_{b}D^{b}-\frac{1}{8}R)\frac{u_1}{r}= \theta_R, 
\]
in  $B_a \backslash \{i\}$, where $\theta_R$ is in $ C^\infty(B_a)$
and vanishes at any order at $i$. 

If $\theta$ has the form (\ref{1f}) then  (\ref{he1}) will be satisfied
due to our assumptions on $h_{ab}$.  

Note the simplicity of condition (\ref{condd}).  If the metric is
analytic on $B_a$ it can be arranged that $\theta_R = 0$ and $u_1$ is
analytic on $B_a$ (and unique with this property, see
\cite{Garabedian}).  The requirement $R > 0$, which ensures the
solvability of the Lichnerowicz equation, could be reformulated in
terms of a condition on the Yamabe number (cf.  \cite{Lee87}). 

Theorem \ref{T1} has two parts. One is the existence and uniqueness of
the solution. This part depends on global properties of the fields on
$S$.  It can be proved under much weaker assumptions on the
differentiability of $h_{ab}$ and $\Psi_{ab}$.  The second part
concerns the regularly in $B_a$, and depends only on local properties
of the fields in $B_a$. This part can also be proved under weaker
hypothesis. These generalizations have  physical relevance, I
will come back to this point in the final section.

It remains to be shown that condition (\ref{condd}) can be satisfied 
by tensor fields $\Psi_{ab}$ which satisfy (\ref{diver}), (\ref{Psii}). A
special class of such solutions, namely those which extend smoothly to all
of $S$, can easily be obtained by known techniques (cf. \cite{Choquet80}).
However, in that case the initial data will have vanishing momentum and
angular momentum. To obtain data without this restriction, we have to
consider fields $\Psi_{ab} \in C^{\infty}(\tilde{S})$ which are singular at
$i$ in the sense that they admit, in accordance with (\ref{pf2}),
(\ref{hPrescale}), (\ref{thetai}), at $i = \{r = 0\}$ asymptotic expansions 
of the form
\begin{equation} 
\label{PsiS}
\Psi_{ij}\sim \sum_{k\geq -4} \Psi^k_{ij} r^k
\quad\mbox{with}\quad \Psi^k_{ij} \in C^{\infty}(S^2). 
\end{equation}

It turns out that condition (\ref{condd}) excludes data with 
non-vanishing linear momentum, which requires a non-vanishing leading order
term of the form $O(r^{-4})$. In section \ref{logs} we will show that such
terms imply terms of the form $\log\,r$ in $\theta$  and thus do not
admit expansion of the form (\ref{he1}). However, this does not
necessarily indicate that condition (\ref{condd}) is overly restrictive. In
the case where the metric $h_{ab}$ is smooth  it will be shown in section
\ref{logs}  that a non-vanishing linear momentum always comes with
logarithmic terms, irrespective of whether condition (\ref{condd}) is
imposed or not.

There remains the question whether there exist fields $\Psi_{ab}$ which
satisfy  (\ref{condd}) and have non-trivial angular momentum. The
latter requires a term of the form $O(r^{-3})$ in (\ref{PsiS}). It turns
out that condition (\ref{condd}) fixes this term to be of the form
\begin{equation} 
\label{PsiJ}
\Psi_{AJ}^{ab}= \frac{A}{r^3}\,(3n^an^b - \delta^{ab})+ \frac{3}{r^3}(n^a  \epsilon^{bcd} J_c n_d +  
n^b\epsilon^{acd} J_c n_d),  
\end{equation}
where $n^i = x^i/r$ is the radial unit normal vector field near $i$ and
$J^k$, $A$  are constants, the three constants $J^k$ specifying the
angular momentum of the data. The spherically symmetric tensor which
appears here with the factor $A$ agrees with the extrinsic curvature for a
maximal (non-time symmetric) slice in the Schwarzschild solution (see for
example \cite{Beig98}). Note that the tensor $\Psi_{AJ}^{ab}$ satisfies
condition (\ref{condd}) and the equation
$\partial_a \Psi^{ab}_{AJ}=0$ on $\tilde S$ for the flat metric, hence 
it is a non-trivial example. We want to study more general situations.

In the existence proof for equations (\ref{diver}) and (\ref{Psii}),
the possible existence of conformal Killing vectors $\xi^a$ of the
metric $h_{ab}$ will play an important role. A conformal Killing
vector is a solution on $S$ of  $\mathcal{L}_h \xi =0$, where we have
defined the conformal Killing operator
\begin{equation}
 \label{eq:calL}
(\mathcal{L}_h v)^{ab} \equiv  D^av^b+D^bv^a - \frac{2}{3}\,h^{ab}\,D_c v^c.
\end{equation}
Given $\xi^a$ we define the followings  constants
\begin{equation}
\label{eq:ckvdata}
S^a={\epsilon^a}_{bc}D^b \xi^c(i), \quad    
a =\frac{1}{3} D_a\xi^a(i). 
\end{equation}

The free-data in the solution $\Psi^{ab}$ consist in two pieces:
a singular and a regular one, which we will denote by  $\Phi^{ab}_{sing}$
and $\Phi^{ab}_{reg}$ respectively.
We define $\Phi^{ab}_{sing}$ in  
$B_a \setminus \{i\}$ by
\begin{equation}
\label{eq:Qn}
\Phi^{ab}_{sing}= \chi\,(\Psi_{AJ}^{ab}-\frac{1}{3}\,
h_{cd}\,\Psi_{AJ}^{cd}\,{h}^{ab}),
\end{equation}
and vanishes elsewhere. Here $\chi$ denotes a smooth function of
compact support in $B_a$ equal to $1$ on $B_{a/2}$.
 We assume that  $\Phi^{ab}_{reg}$   can be written near $i$ in the form
\begin{equation}
\label{eq:conregQ}
\Phi^{ab}_{reg} = r^{-3}\,\Phi^{ab}_{1 reg} + \Phi^{ab}_{2 reg}, 
\end{equation}
where 
$\Phi^{ab}_{1 reg}$, $\Phi^{ab}_{2 reg}$ are smooth symmetric trace
free tensors in $S$
and such that $\Phi^{ij}_{1 reg} = O(r^{2})$,
and $x_i\,x_j\,\Phi^{ij}_{1 reg} = r^2\,\Phi$ with some 
$\Phi \in C^{\infty}(B_a)$.

\begin{theorem} 
\label{T2}
Let $h_{ab}$  a smooth metric in $S$. Let  $\Phi^{ab}_{sing}$
and $\Phi^{ab}_{reg}$ given by (\ref{eq:Qn}) and (\ref{eq:conregQ}). 

i) If the metric $h$ admits no conformal Killing fields on $S$, then
there exists a unique vector field $v^a$  such that the tensor field
\begin{equation}
\label{eq:singe}
\Psi^{ab} = \Phi_{sing}^{ab} + \Phi_{reg}^{ab} 
+ (\mathcal{L}_{h} v)^{ab},
\end{equation}
satisfies the equation $D_a \Psi^{ab}=0$ in $S\setminus \{i\}$. The
solution satisfies $\Psi_{ab} = O(r^{-3})$ at $i$  and  condition
(\ref{condd}).\\ 

\noindent
ii) If the metric $h$ admits conformal Killing fields $\xi^a$ on $S$,
a vector field $v^a$ as specified above exists if and only if the
constants  $J^a$, $A$  (partly) characterizing the
tensor field $\Phi_{sing}^{ab}$  satisfy the equation
\begin{equation}
\label{eq:sycond}
 J^a\,S_a + A\,a  = 0,    
\end{equation}
for any conformal Killing field $\xi^a$ of $h$, where the constants 
 $S_a$ and  $a$ are given  by 
(\ref{eq:ckvdata}).\\    

In both cases the angular momentum of $\Psi^{ab}$ is given by $J^a$. 
 This quantity can thus be prescribed
freely in case (i).
\end{theorem}

Theorem \ref{T2}, as Theorem \ref{T1}, contains two parts. One is the
existence and uniqueness of the solution $v^a$. The restriction
(\ref{eq:sycond}) appears in this part. See \cite{Beig96},
\cite{Dain99} and \cite{Dain01d} for a generalization and
interpretation of this condition.
The other part concerns the 
regularity of the solution in $B_a$, namely that it satisfies condition 
(\ref{condd}).  This part depends only on local assumptions of the
fields in $B_a$.
We prove a more detailed version of this theorem in \cite{Dain99}.

\section{Solution of the Hamiltonian Constraint with Logarithmic 
terms} 
\label{logs}

Assume, for simplicity, that the metric $h_{ab}$ is conformally flat
in $B_a$. Then the operator that appears in the left hand side of
equation \eqref{Lich} reduce to the Laplacian $\Delta$. This
simplification is of course a restriction on the allowed initial data,
but it already contains all the main problems and the essential
features of the more general case. It is also important to remark that
even under this assumption it is possible to describe a rich family of
initial data, since we are making restrictions only on $B_a$.
Consider the Poisson equation \eqref{eq:2}. 
The example 
\begin{equation} \label{logg}
\Delta(\log r  h_m)=r^{-2} h_m (2m+1),
\end{equation}
where $h_m$ is an harmonic polynomial of order $m$ (i.e. $\Delta
h_m=0$) shows that logarithmic terms can occur in solutions to the
Poisson equation even if the source has only terms of the form $r^sp$
where $p$ is some polynomial.  That is, even if our free data have
expansion in powers of $r$ logarithmic terms will  appear in
the conformal factor $\theta$.  We shall use this to show that
logarithmic terms can occur in the solution to the Lichnerowicz
equation if the condition $r^8\Psi_{ab}\Psi^{ab} \in E^\infty (B_a)$
is not satisfied.  Our example will be concerned with initial data
with non-vanishing linear momentum.

We assume that in a small ball $B_a$ centered at $i$
the tensor $\Psi^{ab}$ has the form 
\begin{equation}
\label{eq:pcr}
\Psi^{ab}=\Psi_P^{ab}+\Psi_R^{ab},
\end{equation}
where $\Psi_P^{ab}$ is given in normal coordinates by 
\begin{equation}
\label{eq:PsiPpb}
\Psi_P^{ab}=\frac{3}{2r^4}  \left( -P^a n^b-P^bn^a 
-(\delta^{ab}-5n^an^b)\,P^cn_c \right),
\end{equation}
and $\Psi_R^{ab}=O(r^{-3})$ is a tensor field such that
$\Psi^{ab}$ satisfies equation (\ref{diver}). We will assume also that 
$\Psi_R^{ab}$ satisfies some mild smoothness condition (cf. \cite{Dain99}).  

 Since $\Psi_P^{ab}$ is trace-free and
divergence-free with respect to the flat metric, we could, of course,
choose $h_{ab}$ to be the flat metric and $\Psi_R^{ab}=0$. This would
provide one of the conformally flat initial data sets discussed in
\cite{York}. We are interested in a more general situation.

\begin{lemma}
  Let $h_{ab}$ be a smooth metric, and let $\Psi^{ab}$ be given by
  (\ref{eq:pcr}).  Then, there exists a unique, positive, solution to
  the Hamiltonian constraint (\ref{Lich}). In $B_a$ it has the form
\begin{multline} 
\label{eq:logexp}
\theta= \frac{w_1}{r} + \frac{1}{2}m 
+ \frac{1}{32}\,r\,\left((9\,(P^i n_i)^2 - 33\,P^2 \right) \\
+ \frac{7}{16}\,m\,r^2\,\left(\frac{5}{4}\,P^2 
+ \frac{3}{5}\,(3\,(P^i n_i)^2-P^2)\,\log r \right) + u_R,
\end{multline}
where $P^iP_i=P^2$, the constant $m$ is the total mass of the initial
data, $w_1$ is a smooth function with $w_1=1 + O(r^2)$, and $u_R\in
C^{2,\alpha}(B_a)$ with $u_R(0)=0$.
\end{lemma}

Since $w_1$ is smooth and $u_R$ is in $C^{2,\alpha}(B_a)$, there
cannot occur a cancellation of logarithmic terms. For non-trivial
data, for which $m \neq 0$, the logarithmic term will always appear.
In the case where $h_{ab}$ is flat (in this case we have $w_1=1$) and
$\Psi_R^{ab}=0$ an expansion similar to (\ref{eq:logexp}) has been
calculated in \cite{Gleiser99}.

\section{Explicit solutions of the momentum constraint}
Instead of given the proof of Theorem \ref{T2}, in this section I want
to present some explicit solutions of equation (\ref{diver}) for
conformally flat and also for axially symmetric metrics. In the first
case we will construct all the solutions, in the second one only some
of them. We will show how to achieve condition (\ref{condd}) in terms
of the free data.  In both cases the solution is constructed in terms
of derivatives of some free functions. That makes them suitable for
explicit computations. In particular, if we assume that these free
functions have compact support, then $\Psi^{ab}$ will also have 
compact  support. We note incidentally  that, as an application, one can
easily construct regular hyperboloidal initial data with non-trivial
extrinsic curvature.

\subsection{The momentum constraint on Euclidean space}

In the following we shall give an explicit constructing of the
smooth solutions to the equation $\partial_a\Psi^{ab}=0$ on the 
3-dimensional Euclidean space $\mathbb{E}^3$ (in suitable coordinates 
$\mathbb{R}^3$ endowed with the flat standard metric) or open subsets of
it. Another method to obtain such solutions has been described in
\cite{Beig96'}, multipole expansions of such tensors have been studied in
\cite{Beig96}. 

Let $i$ be a point of $\mathbb{E}^3$ and $x^k$ a Cartesian coordinate
system with origin $i$ such that in these coordinates the metric of
$\mathbb{E}^3$, denoted by $\delta_{ab}$, is given by the standard
form $\delta_{kl}$. We denote by $n^a$ the vector field on 
$\mathbb{E}^3 \setminus \{i\}$ which is given in these coordinates by
$x^k/|x|$.

Denote by $m_a$ and its complex conjugate $\bar{m}_a$ complex vector
fields, defined on $\mathbb{E}^3$ outside a lower dimensional subset and
independent of $r = |x|$, such that 
\begin{equation} 
\label{nm}
m_a m^a=\bar{m}_a\bar{m}^a=n_am^a=n_a\bar{m}^a=0,\,\,\,\,\,\,
m_a \bar{m}^a=1.
\end{equation}
There remains the freedom to perform rotations
$m_a \rightarrow e^{if} m_a$
with functions $f$ which are independent of $r$. 

The metric has the expansion
\[
\delta_{ab}=n_an_b+\bar m_a m_b+ m_a\bar m_b,
\]
while an arbitrary symmetric, trace-free tensor $\Psi_{ab}$ can be
expended in the form 
\begin{multline} 
\label{Psit}
r^{3}\Psi_{ab}= \xi(3n_a n_b -\delta_{ab})+\sqrt{2}\eta_1 
n_{(a} \bar m_{b)}+ \sqrt{2}\bar \eta_1 n_{(a} m_{b)}+ \\
\bar \mu_2 m_a 
m_b+\mu_2 \bar m_a \bar  m_b,
\end{multline}
with
\[
\xi = \frac{1}{2}r^3 \Psi_{ab} n^a n^b, \quad
\eta_1 = \sqrt{2} r^3 \Psi_{ab} n^a m^b, \quad
\mu_2 = r^3 \Psi_{ab} m^a m^b.
\]
Since  $\Psi_{ab}$ is real, the function $\xi$ is real while $\eta_1$, 
$\mu_2$ are complex functions of spin weight 1 and 2 respectively.

Using in the equation
\begin{equation} 
\label{divp}  
\partial_a \Psi^{ab}=0, 
\end{equation}
the expansion (\ref{Psit}) and contracting suitably with $n^a$ and 
$m^a$, we obtain the following representation of
(\ref{divp})   
\begin{equation} 
\label{n1}
4r\partial_r \xi + \bar \eth \eta_1 +\eth \bar \eta_1=0,
\end{equation}
\begin{equation} 
\label{q1}
r\partial_r \eta_1 + \bar \eth \mu_2 - \eth \xi =0.
\end{equation}
Here $\partial_r$ denotes the radial derivative and $\eth$ the edth
operator of the unit two-sphere (cf. \cite{Penrose} for
definition and properties). By our assumptions the differential operator
$\eth$ commutes with $\partial_r$. 

Let ${}_sY_{lm}$ denote the spin weighted spherical harmonics, which
coincide with the standard spherical harmonics $Y_{lm}$ for $s=0$. The
${}_sY_{lm}$ are  eigenfunctions of the operator $\bar \eth \eth$ for
each spin weight $s$.
More generally, we have
\begin{equation}
\label{eq:betheth2}
\bar \eth^p \eth^p ({}_sY_{lm})=(-1)^p \frac{(l-s)!}{(l-s-p)!}\frac{(l+s+p)!}{(l+s)!} {}_sY_{lm}.
\end{equation}

If $\mu_s$ denotes a smooth function on the two-sphere of integral
spin weight $s$, there exists a function $\mu$ of spin weight zero such
that 
$\eta_s=\eth^s \eta$. We set
$\eta^R=\textnormal{Re}(\eta)$ and
$\eta^I=i\,\textnormal{Im}(\eta)$, such that 
$\eta= \eta^R + \eta^I$, and define 
\[
\eta_s^R=\eth^s \eta^R, \,\,\, \eta_s^I=\eth^s \eta^I,
\]
such that $\eta_s= \eta_s^R +\eta_s^I$. We have
\[
\overline{\eth^s\eta^R} = \bar\eth^s \eta^R, \,\,\, 
\overline{\eth^s\eta^I} = - \bar\eth^s \eta^I.
\]

Using these decompositions now for $\eta_1$ and $\mu_2$, we obtain 
equation (\ref{n1}) in the form 
\begin{equation} 
\label{e1}
2r\partial_r \xi =- \bar  \eth  \eth \eta^R.
\end{equation}
Applying $\bar \eth$ to both sides of equation (\ref{q1}) and
decomposing into real and imaginary part yields 
\begin{equation} 
\label{e2}
r\partial_r \bar \eth \eth \eta^I=- \bar{\eth}^2\eth^2  \mu^I
\end{equation}
\begin{equation} \label{e3}
2r\partial_r(r\partial_r \xi) 
+ \bar \eth \eth \xi =\bar{\eth}^2\eth^2 \mu^R.
\end{equation}

Since the right hand side of (\ref{e1}) has an expansion in spherical
harmonics with $l\geq 1$ and the right hand sides of (\ref{e2}),
(\ref{e3}) have expansions with $l\geq 2$, we can determine the expansion
coefficients of the unknowns for $l=0,1$. They can be given in the form
\[
\xi=A+rQ +\frac{1}{r} P,\quad
\eta^I=iJ + const.,\quad
\eta^R=rQ-\frac{1}{r} P + const.,
\]
with 
\begin{equation}
\label{eq:scon}
P=\frac{3}{2} P^an_a,  
\quad Q=\frac{3}{2}Q^an_a, 
\quad  J=3J^an_a,
\end{equation}
where $A, P^a, Q^a, J^a$ are arbitrary constants. 
Using (\ref{Psit}), we obtain the corresponding tensors in the form
(cf. (\cite{York})
\begin{align}
\label{eq:PsiPp}
\Psi_P^{ab}&=\frac{3}{2r^4}  \left( -P^a n^b-P^bn^a 
-(\delta^{ab}-5n^an^b)\,P^cn_c \right),\\
\label{eq:PsiJp}
\Psi_J^{ab}&=\frac{3}{r^3}(n^a  \epsilon^{bcd} J_c n_d +  
n^b\epsilon^{acd} J_c n_d),  \\
\label{eq:PsiAp}
\Psi_A^{ab}&=\frac{A}{r^3}\,(3n^an^b - \delta^{ab}),\\
\label{eq:PsiQp}
\Psi_Q^{ab}&=\frac{3}{2r^2}  \left( Q^a n^b+Q^bn^a 
- (\delta^{ab}-n^an^b)\,Q^cn_c \right). 
\end{align}

We assume now that $\xi$ and $\eta^I$ have expansions in terms of in
spherical harmonics with $l\geq 2$. Then there exists a smooth function
$\lambda_2$ of spin weight 2 such that
\[
\xi=\bar\eth^2 \lambda_2^R, \quad 
\eta_1^I=\bar \eth \lambda_2^I.
\] 
Using these expressions in equations (\ref{e1}) -- (\ref{e3})
and observing that for smooth spin weighted functions $\mu_s$ with $s>0$
we can have $\bar \eth \mu_s=0$ only if $\mu_s=0$, we obtain
\[
\eth \eta^R=-2r\partial_r\bar\eth \lambda_2^R, \quad
\eth^2 \mu^I=-r\partial_r \lambda_2^I,
\]
\[
\eth^2 \mu^R=2r\partial_r(r\partial_r\lambda_2^R)
-2\lambda_2^R+\eth \bar\eth\lambda_2^R.
\]
We are thus in a position to describe the general form of the
coefficients in the expression (\ref{Psit}) 
\begin{align}
\label{eq:xi}
\xi &= \bar\eth^2 \lambda_2^R + A + r\,Q + \frac{1}{r}\,P,\\
\label{eq:eta1}
\eta_1 &= -2\,r\,\partial_r\,\bar\eth \lambda_2^R
+ \bar\eth \lambda_2^I
+ r\,\eth Q - \frac{1}{r}\,\eth P 
+ i\,\eth J,\\
\label{eq:mu2}
\mu_2 &= 2\,r\,\partial_r(r\,\partial_r\,\lambda_2^R)-2\,\lambda_2^R
+ \eth \bar\eth\lambda_2^R-r\,\partial_r\,\lambda_2^I.
\end{align}

\begin{theorem} 
\label{t2}
Let $\lambda$ be an arbitrary complex $C^{\infty}$ function in 
$B_a \setminus \{i\} \subset \mathbb{E}^3$ with 
$0 < a \le \infty$, and set $\lambda_2=\eth^2\lambda$. 
Then the tensor field
\begin{equation}
\label{gensol}
\Psi^{ab} = \Psi_P^{ab} + \Psi_J^{ab}
+ \Psi_A^{ab} + \Psi_Q^{ab} + \Psi_{\lambda}^{ab},
\end{equation}
where the first four terms on the right hand side are given by
(\ref{eq:PsiPp}) -- (\ref{eq:PsiAp}) while $\Psi_{\lambda}^{ab}$ 
is is obtained by using in (\ref{Psit}) only the part of the
coefficients (\ref{eq:xi}) -- (\ref{eq:mu2}) which depends on
$\lambda_2$, satisfies the equation $D^a\Psi_{ab}=0$ in 
$B_a \setminus \{i\}$. 
Conversely, any smooth solution in $B_a \setminus \{i\}$ of
this equation is of the form (\ref{gensol}).
\end{theorem} 

Obviously, the smoothness requirement on $\lambda$ can be relaxed since
$\Psi_{\lambda}^{ab} \in C^1(B_a \setminus \{i\})$ if
$\lambda \in C^5(B_a \setminus \{i\})$. Notice, that no fall-off behavior
has been imposed on $\lambda$ at $i$ and that it can show all kinds of
bad behavior as $r \rightarrow 0$. 

Since we are free to choose the radius $a$, we also obtain an expression
for the general smooth solution on $\mathbb{E}^3 \setminus \{i\}$. By
suitable choices of $\lambda$ we can construct solutions
$\Psi_{\lambda}^{ab}$ which are smooth on $\mathbb{E}^3$ or which are
smooth with compact support.  
Finally we  provide tensor fields which satisfy
condition (\ref{condd}) and thus prove a special case of theorem
\ref{T2}, see \cite{Dain99} for the  proof.

\begin{theorem} 
\label{flatang}
Denote by $\Psi^{ab}$ a tensor field of the type (\ref{gensol}).
If $r\lambda \in  E^\infty(B_a)$ and $P^a=0$, then  
$r^8\Psi_{ab}\Psi^{ab} \in E^\infty(B_a)$.
\end{theorem}

We wish to point out a further application of the results above. 
Given a subset S of $\mathbb{R}^3$ which is compact with boundary, we can
use the representation (\ref{gensol}) to construct hyperboloidal initial
data (\cite{Friedrich83}) on $S$ with a metric $h$ which is Euclidean on
all of $S$ or on a subset $U$ of S. In the latter case we would require
$\Psi_{\lambda}^{ab}$ to vanish on $S \setminus U$. In the case where the
trace-free part of the second fundamental form implied by $h$ on
$\partial S$ vanishes and the support of $\Psi^{ab}$ has empty
intersection with $\partial S$ the smoothness of the corresponding
hyperboloidal initial data near the boundary follows from the discussion
in (\cite{Andersson92}). Appropriate requirements on
$h$ and $\Psi^{ab}$ near $\partial S$ which ensure the smoothness of the
hyperboloidal data under more general assumptions can be found in
(\cite{Andersson94}).

\subsection{Axially symmetric initial data}
 
The momentum constraint with axial symmetry has been studied in
\cite{Baker99b}, \cite{Brandt94a} and \cite{Dain99b}.  Assume that the
metric $h_{ab}$ has a Killing vector $\eta^a$, which is hypersurface
orthogonal. We define $\eta$ by $\eta=\eta^a \eta^b h_{ab}$. Following
\cite{Hawking73b}, consider  $\Psi^{ab}$  defined by
\begin{equation}
  \label{eq:axialpsi}
  \Psi^{ab}=\frac{2\Psi^{(a} \eta^{b)}}{\eta} ,
\end{equation}
where $\Psi^a$ satisfies
\begin{equation}
 \label{eq:J}
\pounds_\eta \Psi^a=0, \quad \Psi^a\eta_a=0, \quad D_a \Psi^a=0,
\end{equation}
with  the Lie derivative  $\pounds_\eta$ with respect $\eta^a$. 

We use the Killing equation $D_{(a}\eta_{b)}=0$,  the fact that
$\eta^a$ is hypersurface orthogonal, (i.e.; it satisfies $D_a
\eta_b=-\eta_{[a}D_{b]} \ln \eta$) and equations (\ref{eq:J})  to
conclude that $\Psi^{ab}$ is trace free and divergence free with respect
to the metric $h_{ab}$.

The solution of equations  (\ref{eq:J}) can be
written in terms of a scalar potential $\omega$
\begin{equation}
  \label{eq:axialve}
  \Psi^a=\frac{1}{\eta} \epsilon^{abc} \eta_b D_c \omega, \quad
  \pounds_\eta \omega =0. 
\end{equation}
Using this equation we find that
\begin{equation}
  \label{eq:11}
  \Psi^{ab}\Psi_{ab}=2\frac{D_c\omega D^c\omega}{\eta^2}.
\end{equation}
We want to find now which conditions we have
to impose in $\omega$ in order to achieve \eqref{condd}.   
The metric $h_{ab}$ has the form
\begin{equation}
  \label{eq:13}
  h_{ab}=e_{ab}+\frac{\eta_a\eta_b}{\eta},
\end{equation}
where $e_{ab}$ is the two dimensional metric induced on the
hypersurfaces orthogonal to $\eta^a$, its satisfies $\pounds_\eta
e_{ab}=0$.  All two dimensional metrics are locally conformally flat,
we will assume here that $e_{ab}$ is globally conformally flat. Then,
we can perform a conformal rescaling of $h_{ab}$ such that in the
rescaled metric the corresponding intrinsic metric $e_{ab}$ is flat.
We will denote this rescaled metric again by $h_{ab}$.  Assume that
$\eta^a$ is a rotation. Take spherical coordinates $(r, \vartheta,
\phi)$ such that $\eta^a=(\partial/\partial\phi)^a$.  In this
coordinates the metric has the form
\begin{equation}
  \label{eq:12}
  h = dr^2+r^2 d\vartheta^2+\eta  d\phi^2.
\end{equation}
The norm $\eta$ can be written as 
\begin{equation}
  \label{eq:14}
  \eta=r^2\sin^2 \vartheta (1+f),
\end{equation}
where $f$  satisfies $1+f>0$. Note that $r$
is the geodesic distance with respect to the origin.

In these coordinates equation \eqref{eq:11} has the form
\begin{equation}
  \label{eq:15}
  \Psi^{ab}\Psi_{ab}=\frac{2}{(1+f)^2}\frac{\left(r^2(\partial_r\omega)^2+
    (\partial_\vartheta \omega)^2\right)}{r^6
    \sin^4 \vartheta }. 
\end{equation}

In the flat case (i.e. $f=0$)  this solution reduce to the one
given by Theorem \ref{t2}  with
\begin{equation}
  \label{eq:9}
  P=Q=A=0,\quad \lambda_2^R=0,
\end{equation}
and
\begin{equation}
  \label{eq:10}
 i \omega= \lambda_2^I \sin^2\vartheta +iJ^z (-3\cos
 \vartheta+\cos^3 \vartheta),
\end{equation}
where $J^z$ is the only non vanished component of the vector $J^a$,
and $\pounds_\eta \lambda_2^I=0$.

Motivated by this expression we write $\omega$ in the form
\begin{equation}
  \label{eq:16}
  \omega= \eth^2\lambda \sin^2\vartheta  +J^z (-3\cos
 \vartheta+\cos^3 \vartheta),
\end{equation}
where $\lambda$ is an arbitrary zero spin real function   which
depends on $r$ and $\vartheta$.  The constant $J^z$ will give the
angular momentum of the data. This can be seen from the following
expression
\begin{equation}
  \label{eq:130}
  J^z= \frac{1}{8\pi} \int_\Sigma \Psi_a  n^a \, dS,
\end{equation}
where $\Sigma$ is any closed two-surface in the asymptotic region and
$n^a$ its normal.  

Using equation \eqref{eq:16} and lemma 9 of \cite{Dain99} we prove
the following lemma.

\begin{theorem}
\label{axialang}
Assume that $f\in E^\infty(B_a)$.
Let  $\Psi^{ab}$ be given by \eqref{eq:axialpsi}, \eqref{eq:axialve}
and \eqref{eq:16}, where $J^z$ is an arbitrary constant.    
If $r\lambda \in  E^\infty(B_a)$  then  
$r^8\Psi_{ab}\Psi^{ab} \in E^\infty(B_a)$.
\end{theorem}

Note that in Theorem \ref{axialang} we have assumed only $f\in
E^\infty(B_a)$, that is $f$, and hence the conformal metric $h_{ab}$,
is not required to be smooth.  I will come back to this point in the
final section.

\section{Main ideas in the proof of theorem \ref{T1}} \label{mainideas}
I want to describe in this section the main idea in the proof of
theorem \ref{T1}. I will concentrate on the regularity part of this
theorem and not on the existence part,  since the later is more or less
standard. I will   give an almost self contained proof of a
simplified version of theorem \ref{T1}, which contains all the
essential elements of the general proof. 

Consider the
semi-linear equation 
\begin{equation}
  \label{eq:5}
\Delta u =f(u,x),  
\end{equation}
on $B_a$, where the function $f$ is given by 
\begin{equation}
  \label{eq:7}
f(x,u)=-\frac{\Psi^2(x)}{(1+ru)^7},  \quad  \Psi(x)\in E^\infty.
\end{equation}
This equation is similar to equation (\ref{Lich}) when the metric
$h_{ab}$ is flat in $B_a$. Assume that we have a positive solution
$u\in C^{2,\alpha}(B_a)$. There exist several method to prove existence
of solutions for semi-linear equations, see for example
\cite{McOwen96} for an elementary introduction to the subject. We want 
to prove the following theorem.

\begin{theorem} \label{T1s}
  If $u\in C^{2,\alpha}(B_a)$ and $u\geq 0$ is a solution of equation
  \eqref{eq:5}, then $u\in E^{\infty}(B_a)$.
\end{theorem}

The important feature of equation \eqref{eq:5} is that the radial
coordinate $r$ appears explicitly in $f$. As a function of the
Cartesian coordinates $x^i$, $r$ is only in $C^{\alpha}$. Thus, we can
not use standard elliptic estimates in order to improve the regularity
of our solution.  Instead of this we use the following spaces.

\begin{definition} 
\label{dEm}
For $m\in \mathbb{N}_0$ and $0<\alpha <1$, we define the space 
$E^{m,\alpha}(B_a)$ as the set
$E^{m,\alpha}(B_a)= \{ f=f_1+rf_2  \, : \,  f_1,f_2  \in C^{m,\alpha}(B_a) \}$.
Furthermore we set 
$E^{\infty}(B_a)= \{ f=f_1+rf_2  \, : \,  f_1,f_2  \in C^{\infty}(B_a) \}$.
\end{definition}

The  spaces $E^{m,\alpha}$ has two important properties. The first one
is given by the following lemma which is an easy consequence of
definition \ref{dEm}.
\begin{lemma} \label{comp}
For $f, g \in E^{m,\alpha}(B_a)$ we have

(i) $f+g\in E^{m,\alpha}(B_a)$ 

(ii)  $fg  \in  E^{m,\alpha}(B_a) $

(iii) If  $f \neq 0$ in $B_a$, then $1/f \in  E^{m,\alpha}(B_a)$.

Analogous results hold for functions in $E^{\infty}(B_a)$.  
\end{lemma}

The second important property  is related
to  elliptic operators. Let $u$ be a solution of the
Poisson equation
\begin{equation}
  \label{eq:1}
  \Delta u =f.
\end{equation}
Then we have the following lemma. 
\begin{lemma}\label{pem}
$f\in E^{m,\alpha}(B_a)  \Rightarrow u\in E^{m+2,\alpha}(B_a)$.
\end{lemma}

{\bf Proof:} Since $f_2 \in C^{m,\alpha}(B_a)$ we can define the
corresponding Taylor polynomial $T_m$ of order $m$. Define $f^R_2$ by
$f_2=T_m+f^R_2$.  It can be seen that $f^R_2=O(r^{m+\alpha})$.  By
explicit calculation we can prove that there exist a polynomial $p_m$
of order $m$ such that $\Delta(r^3p_m)=rT_m$.  Set $u=r^3p_m +u_R$.
Then $u_R$ satisfies the equation
\begin{equation}
  \label{eq:4}
\Delta u_R=f_1+rf^R_2.  
\end{equation}
One can prove that $rf^R_2 \in C^{m,\alpha}(B_a)$.  This is
not trivial because $r$ is only in $C^{\alpha}(B_a)$. Here we use that
$f^R_2=O(r^{m+\alpha})$ (see lemma 3.6 in \cite{Dain99}).  Then the
right-hand side of equation \eqref{eq:4} is in $C^{m,\alpha}(B_a)$.
By the standard Schauder elliptic regularity (see \cite{Gilbarg}) we
conclude that $u_R \in C^{m+2,\alpha}(B_a)$. Thus $u\in
E^{m+2,\alpha}(B_a)$.
\qed\\

As an application of lemmas \ref{comp} and \ref{pem} we can prove
theorem \ref{T1s}. 
Using lemma \ref{comp}, we have that the  function $f$ satisfies
 satisfies the following property
\begin{equation}
  \label{eq:6}
u\in E^{m,\alpha}(B_a) \Rightarrow f(u(x),x) \in E^{m,\alpha}(B_a),
\end{equation}
for every positive $u$.  Assume that we have a solution $u\in
C^{2,\alpha}$ of equation \eqref{eq:5}.  Using \eqref{eq:6}, lemma
\ref{pem} and induction in $m$ we conclude that $u\in E^\infty(B_a)$.
Here we have used  that if $f\in E^{m,\alpha}(B_a)$ for all
$m\in \mathbb{N}_0$, then $f\in E^\infty(B_a)$,   see lemma 3.8 in
\cite{Dain99}.

Note that the only property of $f$ that we have used in order to prove theorem
\ref{T1s} is (\ref{eq:6}).   
Lemma \ref{pem}  can be generalized for second order elliptic
operators with smooth coefficients (cf. \cite{Dain99}). 

\section{Final Comments}

In this section I want to make some remarks concerning the
differentiability of  the initial data.  It is physically
reasonable to assume that the fields are smooth in the asymptotic
region $\tilde S \setminus \Omega$. However in the source region
$\Omega$ smoothness is a restriction. For example, the matter density
of an star is discontinuous at the boundary.  Generalizations of the
existence theorems to include such situations have been made in
\cite{Choquet99} and \cite{Dain01d}.

It is also a restriction to assume that the conformal metric is smooth
in a neighborhood $B_a$ of the point $i$. The point $i$ is the
infinity of the data, hence there is no reason a priori to assume that
the fields there will have the same smoothness as in the
interior. Moreover it has been proved  in \cite{Dain01b} that the stationary
space times do not have a smooth conformal metric. In this case the metric
$h_{ab}$ in $B_a$ has the following form
\begin{equation}
  \label{eq:8}
  h_{ij}=h^1_{ij}+r^3h^2_{ij},
\end{equation}
where $h^1_{ij}$ and $h^2_{ij}$ are analytic functions of the
coordinates $x^i$.  Theorems \ref{T1} and \ref{T2} are generalized in
\cite{Dain01} for metrics that satisfy (\ref{eq:8}). An example of
this generalization is Theorem \ref{axialang}, in which we have
assumed that $f\in E^\infty$, this assumption allows metrics of the
form (\ref{eq:8}).


\begin{thebibliography}{10}

\bibitem{Andersson94}
L.~Andersson and P.~Chru{\'s}ciel.
\newblock On hyperboloidal {C}auchy data for vacuum {E}instein equations and
  obstructions to the smoothness of scri.
\newblock {\em Commun. Math. Phys.}, 161:533--568, 1994.

\bibitem{Andersson92}
L.~Andersson, P.~Chru\'sciel, and H.~Friedrich.
\newblock On the regularity of solutions to the {Y}amabe equation and the
  existence of smooth hyperboloidal initial data for {E}instein's field
  equations.
\newblock {\em Commun. Math. Phys.}, 149:587--612, 1992.

\bibitem{Baker99b}
J.~Baker and R.~Puzio.
\newblock A new method for solving the initial value problem with application
  to multiple black holes.
\newblock {\em Phys. Rev. D}, 59:044030, 1999.

\bibitem{Beig96'}
R.~Beig.
\newblock {TT}-tensors and conformally flat structures on 3-manifolds.
\newblock In P.~Chru{\'s}ciel, editor, {\em Mathematics of Gravitation, Part
  1}, volume~41. Banach Center Publications, Polish Academy of Sciences,
  Institute of Mathematics, Warszawa, 1997.
\newblock gr-qc/9606055.

\bibitem{Beig96}
R.~Beig and N.~O. Murchadha.
\newblock The momentum constraints of general relativity and spatial conformal
  isometries.
\newblock {\em Commun. Math. Phys.}, 176(3):723--738, 1996.

\bibitem{Beig98}
R.~Beig and N.~O. Murchadha.
\newblock Late time behavior of the maximal slicing of the {S}chwarzschild
  black hole.
\newblock {\em Phys. Rev. D}, 57(8):4728--4737, 1998.

\bibitem{Beig91c}
R.~Beig and N.~O'Murchadha.
\newblock Trapped surfaces due to concentration of gravitational radiation.
\newblock {\em Phys. Rev. Lett.}, 66:2421--2424, 1991.

\bibitem{Beig80}
R.~Beig and W.~Simon.
\newblock Proof of a multipole conjecture due to {G}eroch.
\newblock {\em Commun. Math. Phys.}, 78:75--82, 1980.

\bibitem{Beig81}
R.~Beig and W.~Simon.
\newblock On the multipole expansion for stationary space-times.
\newblock {\em Proc. R. Lond. A}, 376:333--341, 1981.

\bibitem{York}
J.~M. Bowen and J.~W. York, Jr.
\newblock Time-asymmetric initial data for black holes and black-hole
  collisions.
\newblock {\em Phys. Rev. D}, 21(8):2047--2055, 1980.

\bibitem{Brandt94a}
S.~Brandt and E.~Seidel.
\newblock The evolution of distorted rotating black holes {III}: Initial data.
\newblock {\em Phys. Rev. D}, 54(2):1403--1416, 1996.

\bibitem{Brill63}
D.~Brill and R.~W. Lindquist.
\newblock Interaction energy in geometrostatics.
\newblock {\em Phys.Rev.}, 131:471--476, 1963.

\bibitem{Choquet99}
Y.~Choquet-Bruhat, J.~Isenberg, and J.~W. York, Jr.
\newblock {E}instein constraint on asymptotically euclidean manifolds.
\newblock {\em Phys. Rev. D}, 61:084034, 1999.
\newblock gr-qc/9906095.

\bibitem{Choquet80}
Y.~Choquet-Bruhat and J.~W. York, Jr.
\newblock The {C}auchy problem.
\newblock In A.~Held, editor, {\em General Relativity and Gravitation},
  volume~1, pages 99--172. Plenum, New York, 1980.

\bibitem{Dain01}
S.~Dain.
\newblock Asymptotically flat initial data with prescribed regularity {II}.
\newblock In preparation., 2001.

\bibitem{Dain99b}
S.~Dain.
\newblock Initial data for a head on collision of two kerr-like black holes
  with close limit.
\newblock {\em Phys. Rev. D}, 64(15):124002, 2001.
\newblock gr-qc/0103030.

\bibitem{Dain01b}
S.~Dain.
\newblock Initial data for stationary space-time near space-like infinity.
\newblock {\em Class. Quantum Grav.}, 18(20):4329--4338, 2001.
\newblock gr-qc/0107018.

\bibitem{Dain99}
S.~Dain and H.~Friedrich.
\newblock Asymptotically flat initial data with prescribed regularity.
\newblock {\em Comm. Math. Phys.}, 222(3):569--609, 2001.
\newblock gr-qc/0102047.

\bibitem{Dain01d}
S.~Dain and G.~Nagy.
\newblock Initial data for fluid bodies in general relativity.
\newblock {\em submitted for publication}, 2001.

\bibitem{Friedrich83}
H.~Friedrich.
\newblock {C}auchy problems for the conformal vacuum field equations in general
  relativity.
\newblock {\em Commun. Math. Phys.}, 91:445--472, 1983.

\bibitem{Friedrich88}
H.~Friedrich.
\newblock On static and radiative space-time.
\newblock {\em Commun. Math. Phys.}, 119:51--73, 1988.

\bibitem{Friedrich98}
H.~Friedrich.
\newblock Gravitational fields near space-like and null infinity.
\newblock {\em J. Geom. Phys.}, 24:83--163, 1998.

\bibitem{Garabedian}
P.~R. Garabedian.
\newblock {\em Partial Differential Equations}.
\newblock John Wiley, New York, 1964.

\bibitem{Geroch70}
R.~Geroch.
\newblock Multipole moments. {II}. curved space.
\newblock {\em J. Math. Phys.}, 11(8):2580--2588, 1970.

\bibitem{Gilbarg}
D.~Gilbarg and N.~S. Trudinger.
\newblock {\em Elliptic Partial Differential Equations of Second Order}.
\newblock Springer-Verlag, Berlin, 1983.

\bibitem{Gleiser99}
R.~J. Gleiser, G.~Khanna, and J.~Pullin.
\newblock Evolving the {B}owen-{Y}ork initial data for boosted black holes.
\newblock {\em gr-qc/9905067}, 1999.

\bibitem{Hawking73b}
S.~W. Hawking.
\newblock The event horizon.
\newblock In C.~DeWitt and B.~S. DeWitt, editors, {\em Black Holes}, pages
  1--56. Gordon and Breach Science Publishers, New York, 1973.

\bibitem{Kundu81}
P.~Kundu.
\newblock On the analyticity of stationary gravitational fields at spatial
  infinity.
\newblock {\em J. Math. Phys.}, 22(9):2006--2011, 1981.

\bibitem{Lee87}
J.~M. Lee and T.~H. Parker.
\newblock The {Y}amabe problem.
\newblock {\em Bull. Amer. Math. Soc.}, 17(1):37--91, 1987.

\bibitem{McOwen96}
R.~C. McOwen.
\newblock {\em Partial Differential Equation}.
\newblock Prentice Hall, New Jersey, 1996.

\bibitem{Misner60}
C.~W. Misner.
\newblock Wormhole initial conditions.
\newblock {\em Phys.Rev.}, 118:1110--1111, 1960.

\bibitem{Misner63}
C.~W. Misner.
\newblock The method of images in geometrostatics.
\newblock {\em Ann. Phys.}, 24:102--117, 1963.

\bibitem{Penrose}
E.~T. Newman and R.~Penrose.
\newblock Note on the {B}ondi-{M}etzner-{S}achs group.
\newblock {\em J. Math. Phys.}, 7(5):863--870, 1966.

\end{thebibliography}

\end{document}